\documentclass[prd,twocolumn,floatfix,preprintnumbers,letterpaper]{revtex4}
\usepackage{amssymb,amsbsy,bm,amsmath,psfrag,dsfont}
\usepackage{hyperref}
\usepackage{graphicx}
\usepackage{epsfig}

\begin{document}

\newcommand{\be}{\begin{equation}}
\newcommand{\ee}{\end{equation}}
\newcommand{\bea}{\begin{eqnarray}}
\newcommand{\eea}{\end{eqnarray}}
\newcommand{\no}{\noindent}
\newcommand{\la}{\lambda}
\newcommand{\si}{\sigma}
\newcommand{\vp}{\mathbf{p}}
\newcommand{\vk}{\vec{k}}
\newcommand{\vx}{\vec{x}}
\newcommand{\om}{\omega}
\newcommand{\Om}{\Omega}
\newcommand{\ga}{\gamma}
\newcommand{\Ga}{\Gamma}
\newcommand{\gaa}{\Gamma_a}
\newcommand{\al}{\alpha}
\newcommand{\ep}{\epsilon}
\newcommand{\app}{\approx}
\newcommand{\nc}{\newcommand}

\title{Instability of Quantum de Sitter Spacetime}
\author{Chiu Man Ho} \email{cmho@msu.edu}
\affiliation{Department of
Physics and Astronomy, Michigan State University, East Lansing, MI 48824, USA}
\author{Stephen~D.~H.~Hsu} \email{hsu@msu.edu}
\affiliation{Department of
Physics and Astronomy, Michigan State University, East Lansing, MI 48824, USA}
\date{\today}

\begin{abstract}

Quantized fields (e.g., the graviton itself) in de Sitter (dS) spacetime lead to particle production: specifically, we consider
a thermal spectrum resulting from the dS (horizon) temperature. The energy required to excite these particles reduces slightly the rate of
expansion and eventually modifies the semiclassical spacetime geometry. The resulting manifold no longer has constant curvature nor time
reversal invariance, and back-reaction renders the classical dS background unstable to perturbations. In the case of AdS, there exists a
global static vacuum state; in this state there is no particle production and the analogous instability does not arise.

\end{abstract}
\maketitle

\section{Introduction}

In classical general relativity, a cosmological constant $\Lambda$ has the special property that
the equation of state must satisfy
\bea
w = \frac{p}{\rho} = -1
\eea
exactly, where $p$ is the associated pressure and $\rho$ the energy density.
This is equivalent to the energy-momentum tensor $T_{\mu \nu}$ satisfying
\bea
T_{\mu \nu} = \Lambda\, g_{\mu \nu} ~~,
\eea
where $g_{\mu \nu}$ is the metric tensor. For $\Lambda > 0$, negative pressure does negative work as the universe expands, and provides exactly
enough energy to produce new spacetime volume
filled with more cosmological constant. Thus, expansion can continue forever, leading to a highly symmetric constant curvature
spacetime known as de Sitter (dS) spacetime.

A $(d+1)$-dimensional dS spacetime is a hyperboloid in a $(d+2)$-dimensional Minkowski spacetime described by
\bea
\label{dS}
x_0^2 - x_1^2 - x_2^2 - \cdots - x_{d+1}^2 =  -3/\Lambda \equiv - R^2 ~~.
\eea
In the global coordinates, the dS metric is
\bea
ds^2 = dt^2 - R^2 \,\cosh^2(t/R)\, d\Omega_d^2 ~~,
\eea
while in the cosmological or Friedmann-Robertson-Walker (FRW) coordinates (which cover only a portion of the global dS manifold (\ref{dS})),
it is given by
\bea
\label{FRW}
ds^2 = dt^2 - a^2(t) \, d\Omega_d^2 ~~,
\eea
where $a(t) = e^{H t}$ with $H =\sqrt{\Lambda/3}$. Our discussion below is focused on the case $d=3$, or four spacetime dimensions.

Quantum excitations, including, inevitably, gravitons (quanta of the gravitational field itself), modify
at least slightly the relation between pressure and energy density. Multi-particle quantum states typically have positive energy density
and pressure, leading to a value of $w$ slightly larger than $-1$, and a pressure insufficiently negative to support dS expansion.

The calculation of quantum contributions to the stress-energy tensor $T_{\mu \nu}$ in dS spacetime is complex, depending on choices of
regularization and vacuum state. For recent progress on this problem, see \cite{Mottola,Mottola1,Parker,Parker1,Parker2,comment}. In this
note, we explore the consequences of the relatively well-understood dS temperature on the macroscopic dynamics of spacetime.

\section{Back-reaction from de Sitter Thermal Spectrum}

In dS spacetime, inertial observers see a thermal distribution of particles and a dS temperature \cite{dS-temperature}.
Observers who detect thermal particles will dispute the notion that the physical
(renormalized) $T_{\mu \nu}$ is proportional to $g_{\mu \nu}$ (i.e., that the equation of state describes pure vacuum energy or
cosmological constant). This deviation from the classical form of $T_{\mu \nu}$ violates dS symmetry. It is due to a subtle infrared
effect (dS temperature) and will not necessarily appear in calculations of UV contributions to the renormalization of $T_{\mu \nu}$.

In \cite{dS-temperature}, Gibbons and Hawking note that detector absorption of thermal radiation from the dS horizon leads, via back-reaction,
to shrinkage of the horizon. We are interested in an averaged, semiclassical $T_{\mu \nu}$ that appears on the right hand side of the Einstein
equations, and results from the steady occurrence of such events throughout spacetime. Interactions will eventually equilibrate each particle
species with the dS horizon temperature.

The fact that inertial observers in dS spacetime see a thermal distribution of particles can also be understood in terms
of the Unruh effect \cite{Unruh}. One can consider dS spacetime as a timelike hyperboloid embedded in a
Minkowski spacetime of one higher (spatial) dimension \cite{dS_Unruh,Deser}. Inertial dS observers, viewed from the perspective of
the embedding spacetime, are uniformly accelerated, and hence their detectors register a thermal bath. From the Unruh perspective, it is
clear that the energy of absorbed thermal particles comes from work done by the accelerating force on the detector \cite{Unruh, Singleton}.
From the dS perspective, this energy comes from work that otherwise would have been performed by the negative pressure. Thus, it clearly
reduces the amount of expansion that would otherwise occur (i.e., in the absence of quantum mechanics) in dS spacetime.

The dS temperature is $T = R^{-1} / 2 \pi$, where $R$ is the dS radius. The ratio of the thermal energy density to cosmological constant is
of order the latter in Planck units, henceforth parametrized by $\epsilon$. The local energy density at late times in the expanding phase of
dS spacetime is therefore slightly larger than in the classical case: $\rho = \Lambda (1 + \epsilon)$. The corresponding pressure is
$p \approx - \Lambda (1 -\xi \, \epsilon)$ where $\xi = 1/3$ and $\xi = 0$ correspond to relativistic and non-relativistic thermal particles
respectively. Thus $w \neq -1$ and the expansion is no longer exactly exponential. However, from the Friedmann equation
\bea
\frac{\ddot{a}}{a} = -\frac{4\pi G}{3}\,\left(\rho+3\,p\right)~~,
\eea
it follows that as long as $\rho >0$ and $w < -1/3$, acceleration is still positive. Therefore, an accelerating expansion of
dS spacetime is still expected. Using the equation of continuity $\dot{\rho} + 3 \,\frac{\dot{a}}{a}\,(\rho + p) = 0$,
one can show that
\bea
\frac{\dot{\epsilon}}{\epsilon} + 3\,(1+\xi) \,\frac{\dot{a}}{a} = 0~~.
\eea
The above equation can be solved to give
\bea
\epsilon \sim a^{-3(1+\xi)}~~.
\eea
In fact, the situation is more complicated than suggested by the simple equations above. As particles produced by earlier expansion are redshifted away, new particles are produced. After many Hubble timescales, the average energy density due to quantum effects should be approximately that of a thermal bath at the dS temperature. (The Bunch-Davies vacuum \cite{BD},
or an approximate version of it, is an attractor.)

Conservation of energy implies that the resulting proper volume of the universe $V$ is slightly smaller than in the classical case:
\begin{equation}
V \approx V_{\textrm{classical}} \, \cdot \, (1 - \epsilon) = \exp (3 H t) \,\cdot \, (1 - \epsilon) ~~,
\end{equation}
and so
\begin{equation}
{ \log V  \over 3t } \approx  H - \epsilon / 3t~~.
\end{equation}
Thus, the spacetime which results from incorporating back-reaction of these quantum effects is no longer one of constant curvature. At late
times, the classical and quantum spacetimes differ macroscopically, despite the smallness of the dS temperature. Expansions about the original
(classical) dS spacetime should exhibit IR instabilities, since dS is not an exact solution once back-reaction is taken into account. Earlier
work has found evidence of instabilities in dS \cite{instability1,instability2}, although the relation to our results is not clear.

The resulting quantum spacetime also cannot be time-reversal invariant. If the late time thermal particle density were also found at early times,
during the contracting phase of global dS spacetime, the resulting blue-shift of thermal particles would lead to radical departure from the vacuum
Einstein equations. (This point has also been emphasized in \cite{Mottola}.) Therefore, the early and late time geometries, taking into account
quantum effects, cannot be the same.

\section{Anti-de Sitter Spacetime}

Anti-de Sitter (AdS) spacetime is similarly a surface of constant (negative) curvature, satisfying the constraint
(for simplicity we restrict to ${\rm AdS}_4$)
\begin{equation}
T^2 + W^2 - X^2  - Y^2 - Z^2 = -3/\Lambda \equiv R^2
\end{equation}
in five-dimensional Minkowski space with metric
\begin{equation}
 ds^2 =  dT^2 + dW^2 - dX^2 - dY^2 - dZ^2 ~~.
\end{equation}

Some AdS worldlines correspond to uniform acceleration in the embedding space \cite{Deser}, again suggesting the presence of
thermal (Unruh) radiation and modification of the semiclassical geometry. However, AdS differs from dS in an important way:
one can define global static coordinates in AdS,
\begin{eqnarray}
\label{global}
  T & = &  R\,\sqrt{1 + r^2/R^2} \,\cos (t/R)  \nonumber \\
  W & = &  R\,\sqrt{1 + r^2/R^2} \,\sin (t/R) \nonumber  \\
  X & = &  r \,\cos\theta  \nonumber  \\
  Y & = &  r\,\sin\theta \,\cos\phi \nonumber  \\
  Z & = &  r\,\sin\theta \,\sin\phi
\end{eqnarray}
with metric
\begin{equation}
ds^2 = \left(1 + \frac{r^2}{R^2}\right) dt^2 - \left( 1 + \frac{r^2}{R^2} \right)^{-1} dr^2 - r^2 \,d\Omega^2 ~~ .
\end{equation}
Although this metric violates spatial translation invariance, the fact that it is static implies that there is a quantum vacuum
state that is time-independent: it does not exhibit particle production or thermal radiation. For this special choice of vacuum state,
AdS is stable to the dS instability discussed above. This result is a consequence of the existence of a global timelike Killing vector.
Other choices of AdS vacuum state, such as the one appropriate to the ``cosmological''
(non-static) coordinates (covering only a portion of global AdS)
 \begin{eqnarray}
 \label{cosmological}
  T & = &  R\,\cos (t/R) \nonumber \\
  W & = &  R\,\sin (t/R) \,\cosh\chi \nonumber \\
  X & = &  R\,\sin (t/R) \,\sinh\chi \,\cos\theta \nonumber \\
  Y & = &  R \,\sin (t/R) \,\sinh\chi \,\sin\theta \,\cos\phi  \nonumber \\
  Z & = &  R \,\sin (t/R) \,\sinh\chi \,\sin\theta \,\sin\phi
\end{eqnarray}
with metric
\begin{equation}
ds^2 =  dt^2 - R^2 \,\sin^2 (t/R) \,\left[\, d\chi^2 + \sinh^2\chi \, d\Omega^2 \,\right]
\end{equation}
do in fact lead to particle production \cite{AdSradiation} and consequent modification of the spacetime geometry.
The difference between the two cases is the choice of quantum vacuum state. The global vacuum state is defined on a spacelike slice
(e.g., at fixed $t$) in coordinates (\ref{global}), but this covers the range $- \infty < T < \infty$ in the embedding space. In contrast,
a fixed $t$ slice in the cosmological coordinates corresponds to fixed $T$. Therefore, the global vacuum state can impose conditions on past
and future quantum states in the cosmological coordinates, which lead to the cancelation of otherwise expected particle production due to
acceleration. Stability of AdS depends on choice of vacuum state, and the spacelike surface on which it is defined.

\section{Remarks}

The argument presented here is the simplest one we know of that indicates the instability of dS spacetime once quantum effects are considered.
The effect we identified is small, but does break the dS symmetries even in the asymptotic regions of the manifold. We do not exclude the possibility
of more dramatic quantum effects, such as significant decay of the cosmological constant itself \cite{instability1,Brandenberger}.
Note that our effect specifically depends on the back-reaction, via the Einstein equations, of the spacetime geometry to modification of
the equation of state. We do not address the possibility that a quantum field (e.g., massive scalar) propagating on a {\it fixed} dS background
could have some intrinsic instability \cite{instability1,instability2,Marolf}.

\emph{Acknowledgements.}\,\, We thank Yu Nakayama for useful conversations. This work was supported by the Office of the Vice-President for
Research and Graduate Studies at Michigan State University.



\begin{thebibliography}{99}
\bibitem{comment}
Calculations of particle production due to cosmological expansion in \cite{Mottola,Mottola1} use the formalism of \cite{Parker,Parker1,Parker2},
in which a vacuum state defined on an early time slice is shown to contain excitations of the vacuum defined on a later slice.
The vacuum states are defined by the condition that they are annihilated by the destruction operators associated with mode functions
of a particular time slice. This suffices to demonstrate particle production, but it is not clear to us whether these calculations (and related results for the energy momentum tensor) are
sensitive to the dS temperature effects considered in this paper. See \cite{dS-temperature} for some relevant discussion.
\end{thebibliography}
\end{document}